%% file: main.tex
\documentclass[10pt,twocolumn,letterpaper]{article}

\usepackage[pagenumbers]{cvpr} 

\input{preamble}
\definecolor{cvprblue}{rgb}{0.21,0.49,0.74}
\usepackage[pagebackref,breaklinks,colorlinks,allcolors=cvprblue]{hyperref}

\usepackage{makecell}
\usepackage{xspace}
\usepackage{multirow}
\usepackage{array}
\usepackage[most]{tcolorbox}
\usepackage{colortbl} 
\newcolumntype{M}{w{c}{0.8cm}} 
\definecolor{cvprblue}{RGB}{45,108,223} 
\newcommand{\benchname}{$\text{M}^3\text{WME-Bench}$\xspace}

\makeatletter
\let\oldcvpr@maketitle\@maketitle
\def\@maketitle{%
  \oldcvpr@maketitle
  \begin{center}
    \includegraphics[width=1\textwidth]{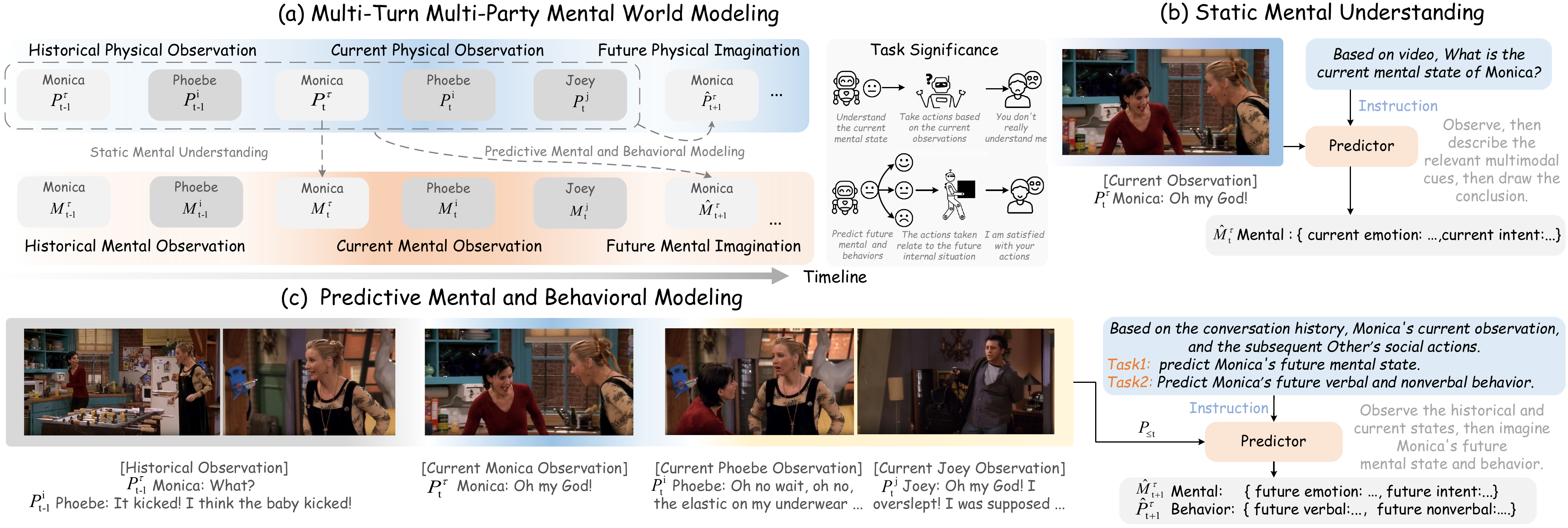}
    \captionof{figure}{\textbf{Tasks illustration.} 
  \textbf{(a) Multi-turn multi-party mental world modeling:} World states factorize into observable physical state $\mathcal{P}$ and latent mental state $\mathcal{M}$, and we benchmark mental modeling, offering insights for planning and interaction. See \Cref{sec:task_formulation} for the formalization.
  Task Comparison: Unlike \textbf{(b) Static Mental Understanding}, which merely infers the target subject's current mental state from an isolated clip, \textbf{(c) Predictive Mental and Behavioral Modeling} requires the model to forecast the target subject's future mental state and behaviors, conditioned on interaction history, current target observations, and observations of others' social actions.}
    \label{fig:task_define}
  \end{center}
}
\makeatother

\def\confName{CVPR}
\def\confYear{2027}

\title{Can You Predict the Next Mind in the Symphony? \\Toward Multi-Turn Multi-Party Mental World Modeling}

\author{
  Jiahao Huang$^{1}$ \quad Yazhou Zhang$^{2}$ \quad Wen Yin$^{1}$ \quad Haoyu Chen$^{3}$ \quad Zheng Lian$^{1}$\thanks{Corresponding Author} \\
  $^1$State Key Laboratory of Autonomous Intelligent Unmanned Systems, Tongji University \\
  $^2$Tianjin University \quad $^3$University of Oulu
}

\begin{document}
\maketitle
\input{sec/0_abstract}    
\input{sec/1_intro}
\input{sec/2_relatedwork}
\input{sec/3_method}
\input{sec/4_experiment}
\input{sec/5_conclusion}

{
    \small
    \bibliographystyle{ieeenat_fullname}
    \bibliography{main}
}

\input{sec/6_suppl}

\end{document}

%% file: sec/0_abstract.tex
\begin{abstract}
Although multimodal foundation models have achieved remarkable progress in modeling, generation, and embodied reasoning in the physical world, their capacity for modeling the dynamic mental world remains underexplored. Existing benchmarks predominantly focus on static mental state perception or single-turn behavioral reasoning, overlooking the dynamic mental transitions of the observed individuals in multi-party, multi-turn interactions. To bridge this gap, we introduce \textbf{\benchname}, a benchmark designed to evaluate mental world modeling conditioned on evolving actions of interactants. Specifically, models are tasked with forecasting a target subject's future mental states and behaviors over extended social interactions. In addition to the benchmark suite, we construct \textbf{M$^3$WME-23K}, a dedicated training dataset, and establish a competitive baseline named \textbf{MindNext-7B}. Extensive evaluations across four representative paradigms (general MLLMs, emotion-specialized MLLMs, agentic frameworks, and World Foundation Models) reveal that current systems consistently struggle to anticipate forward-looking mental states. Mechanistic analyses further reveal that existing architectures fundamentally lack forward mental state transition dynamics. Our findings highlight a pivotal challenge in advancing agents from physical to mental world modeling, charting new directions for socially adaptive AI. 
\end{abstract}

%% file: sec/1_intro.tex
\section{Introduction}
\label{sec:intro}

Physical world models~\cite{assran2025vjepa2,openai2024sora,bruce2024genie,hafner2019planet} enable intelligent agents to internally simulate and forecast environment dynamics, serving as the predictive bedrock for planning and decision-making~\cite{ha2018world,lecun2022path}. Yet, as embodied agents enter shared human environments, effective interaction and collaboration demands go beyond physical dynamics and spatial kinematics~\cite{fei2026mental,ryan2025toward}. In open social contexts, human behavior is fundamentally governed by unobservable mental states, \eg, beliefs, desires, intentions, and emotions. Consequently, agents must transcend physical simulation and advance toward mental world modeling, acquiring the capacity to anticipate dynamic human mental states.

To formalize this cognitive leap, recent studies have pioneered the mental world models~\cite{liu2026modeling,fei2026mental} and the social world model~\cite{zhou2026social}, expanding the latent state space from physical variables to psychological attributes through recursive inference. However, natural social interaction is rarely static or purely dyadic; it inherently unfolds as a multi-party, multi-turn continuum over extended horizons. As conceptualized in \Cref{fig:task_define}~(a), a mental world model must integrate the extended interaction history with observations of the target, while accounting for interactants' social actions as contextual signals that continuously modulate mental states. Furthermore, because latent mental states fundamentally drive outward behavior~\cite{baker2009action,premack1978chimpanzee}, mental modeling cannot remain an isolated attribution exercise. Instead, it must couple internal psychological transitions with future action prediction, thereby grounding unobservable mental states in observable behavioral trajectories and providing the crucial foresight for downstream agent planning.

Existing benchmarks exhibit profound discrepancies when measured against these requirements (\Cref{tab:benchmark_comparison}). As shown in~\Cref{fig:task_define} (b), traditional benchmarks~\cite{busso2008iemocap,poria2019meld,zhang2022mintrec,zhang2024mintrec2} remain largely confined to static mental state perception. By formulating evaluation around post-hoc recognition from observed cues, they merely assess correlation-based attribution, inherently overlooking forward-looking temporal dynamics. Meanwhile, although recent exploratory efforts have begun investigating predictive transitions~\cite{wang2026hief,hu2026emotrans}, next-utterance forecasting~\cite{yang2026saynext}, and multimodal Theory-of-Mind tasks~\cite{shi2025muma,fei2026mental}, they predominantly operate on isolated, turn-level transitions within dyadic exchanges, neglecting the cumulative interaction history and multi-party context shaped by peer actions. Crucially, existing benchmarks decouple mental attribution from behavior, failing to evaluate internal transitions alongside subsequent actions.

\begin{table}[t]
\centering
\caption{Attributes comparison between \benchname and other widely used benchmarks. PMM and PBM represent predictive mental modeling and behavioral modeling respectively.}
\label{tab:benchmark_comparison}
\footnotesize
\setlength{\tabcolsep}{1.7pt}
\renewcommand{\arraystretch}{1.15}
\begin{tabular}{lccccc}
\toprule
  Benchmarks &
  \rotatebox[origin=bc]{37}{Modality} &
  \rotatebox[origin=bc]{37}{Multi-turn} &
  \rotatebox[origin=bc]{37}{Multi-party} &
  \rotatebox[origin=bc]{37}{PMM} &
  \rotatebox[origin=bc]{37}{PBM} \\
\midrule
IEMOCAP~\cite{busso2008iemocap} &Multimodal  & $\checkmark$ & $\times$ & $\times$     & $\times$ \\
MultiWoZ~\cite{budzianowski2018large} &Unimodal & $\checkmark$ & $\times$ & $\times$    & $\times$ \\
MELD~\cite{poria2019meld}    & Multimodal & $\checkmark$ & $\checkmark$ & $\times$     & $\times$ \\
MIntRec~2.0~\cite{zhang2024mintrec2} & Multimodal & $\checkmark$ & $\checkmark$ & $\times$     & $\times$\\
MTMEUR~\cite{hu2025beyond} &Multimodal & $\checkmark$ & $\times$ & $\checkmark$    & $\times$ \\
Hi-EF~\cite{wang2026hief}   & Multimodal & $\times$ & $\times$     & $\checkmark$     & $\times$ \\
Menti-Bench~\cite{fei2026mental} &Multimodal & $\times$ & $\times$     & $\times$     & $\checkmark$ \\
EmoTrans~\cite{hu2026emotrans} &Multimodal & $\times$ & $\checkmark$  & $\checkmark$ & $\times$ \\
EWH~\cite{song2025large}   & Multimodal   & $\times$ & $\checkmark$ & $\checkmark$ & $\times$ \\
EPC~\cite{shahriar2019epc} & Multimodal & $\checkmark$ & $\times$ & $\checkmark$  & $\times$ \\
\midrule
\textbf{Ours} &Multimodal & $\checkmark$ & $\checkmark$ & $\checkmark$ & $\checkmark$ \\
\bottomrule
\end{tabular}
\end{table}

To bridge these gaps, we present \benchname (\textbf{M}ulti-turn \textbf{M}ulti-party \textbf{M}ental \textbf{W}orld \textbf{M}odeling \textbf{E}valuation Benchmark), as conceptualized in \Cref{fig:task_define}(c). Grounded in rich contexts, \benchname conditions evaluation on multi-party interaction histories, instantaneous target observations, and subsequent social cues from surrounding interactants. Our benchmark comprises two tasks: Predictive Mental Modeling (PMM) for the target subject's future mental state, and Predictive Behavioral Modeling (PBM) for future verbal and nonverbal behavior. Spanning diverse social relationships and environments, \benchname provides 3,858 question-answer (QA) pairs under a standardized and reproducible protocol. Furthermore, to facilitate training with strict benchmark isolation, we curate M$^3$WME-23K, a dedicated dataset of 23K QA pairs across 169 interaction hours. Leveraging this data, we train MindNext-7B, an omni-modal language model serving as a strong baseline for future research.

Extensive evaluations across four representative paradigms (general MLLMs, emotion-specialized MLLMs, agentic frameworks, and world foundation models) reveal that existing systems fall short of accurately forecasting future mental states and behaviors. Through subsequent mechanistic analyses, we further illuminate the underlying factors responsible for this limitation. Through \benchname, we aspire to advance the development of mental world models, fostering next-generation affective interaction and socially adaptive human-AI collaboration.

In summary, our primary contributions are as follows:
\begin{itemize}
    \item We propose \textbf{\benchname}, an evaluation benchmark dedicated to assessing dynamic mental world modeling in multi-turn, multi-party interactions. By conditioning on evolving social contexts, it establishes a unified formulation to jointly evaluate prospective mental state transitions and subsequent behavioral actions.
    \item To decouple evaluation from training and facilitate model development, we curate \textbf{M$^3$WME-23K}, a dedicated multimodal training dataset featuring 23K QA pairs across 169 hours of interactions, upon which we train an omni-modal baseline, \textbf{MindNext-7B}.
    \item We conduct comprehensive evaluations on 21 representative models spanning four paradigms. Our empirical and mechanistic analyzes uncover a critical absence of forward mental transition dynamics in prevailing architectures, providing actionable insights for developing socially adaptive artifical intelligence.
\end{itemize}

%% file: sec/2_relatedwork.tex
\section{Related Works}
\label{sec:rw}

\noindent{\textbf{Static Mental Understanding Benchmarks.}} 
A substantial body of benchmarks is dedicated to assessing models' comprehension of instantaneous psychological states. IEMOCAP~\cite{busso2008iemocap} and MELD~\cite{poria2019meld} pioneer conversational emotion recognition in dyadic and multi-party scenarios, respectively, while MIntRec~\cite{zhang2022mintrec, zhang2024mintrec2} and MEIJU~\cite{liu2025meiju} focus on multimodal human intent recognition. Recent endeavors transition from isolated affective attributes toward joint understanding~\cite{huang2026nanoemox}: EmoBench-M~\cite{hu2026emobenchm} benchmarks MLLMs on unified emotion and intent perception, and MC-EIU~\cite{liu2024mceiu} probes cross-lingual emotion-intent coupling. Beyond overt emotional expressions, Theory-of-Mind (ToM) benchmarks such as ToMBench~\cite{chen-etal-2024-tombench}, OpenToM~\cite{xu-etal-2024-opentom}, and MoMentS~\cite{villa-cueva-etal-2025-moments} explore the inference of latent beliefs and mental states. 
Ultimately, these benchmarks remain present-anchored, capturing only instantaneous mental states while overlooking their continuous temporal dynamics.

\noindent{\textbf{Predictive State Modeling Benchmarks.}} 
More recent benchmarks have shifted the focus from retrospective perception to forward-looking anticipation. The early studies by Shahriar and Kim~\cite{shahriar2019epc} laid the empirical groundwork for affective forecasting; Hi-EF~\cite{wang2026hief} introduced the first dedicated dataset for dyadic emotion forecasting, and PEC~\cite{altarawneh2024pec} systematically investigated arousal-driven affective anticipation. Expanding this predictive horizon, SayNext-Bench~\cite{yang2026saynext} forecasts next utterances from multimodal contexts, I2M~\cite{umagami2025i2m} projects intent-driven physical body actions, and emerging affective world models~\cite{song2025large} simulate future visual scenes under emotional and interpersonal influences. 

Despite these advancements, existing benchmarks face two key bottlenecks: evaluating isolated dimensions (mental states, utterances, or actions) rather than their unified cognitive dynamics, and relying on single-turn contexts that lack structured episodic memory and peer causal interventions.
To transcend these limitations, we introduce \benchname, a holistic benchmark tailored for multi-turn, multi-party mental world modeling. 

%% file: sec/3_method.tex
\section{\texorpdfstring{\benchname}{M3WME-Bench}}
This section presents the overarching formulation and construction of our benchmark. We first mathematically formulate the task of predictive mental-behavioral modeling. Next, we elaborate on the data curation pipeline for \benchname and the {M$^3$WME-23K} training suite, followed by key dataset statistics and the implementation of our strong omni-modal baseline, MindNext-7B.
\begin{figure*}[t]
    \centering
    \includegraphics[width=1.0\linewidth]{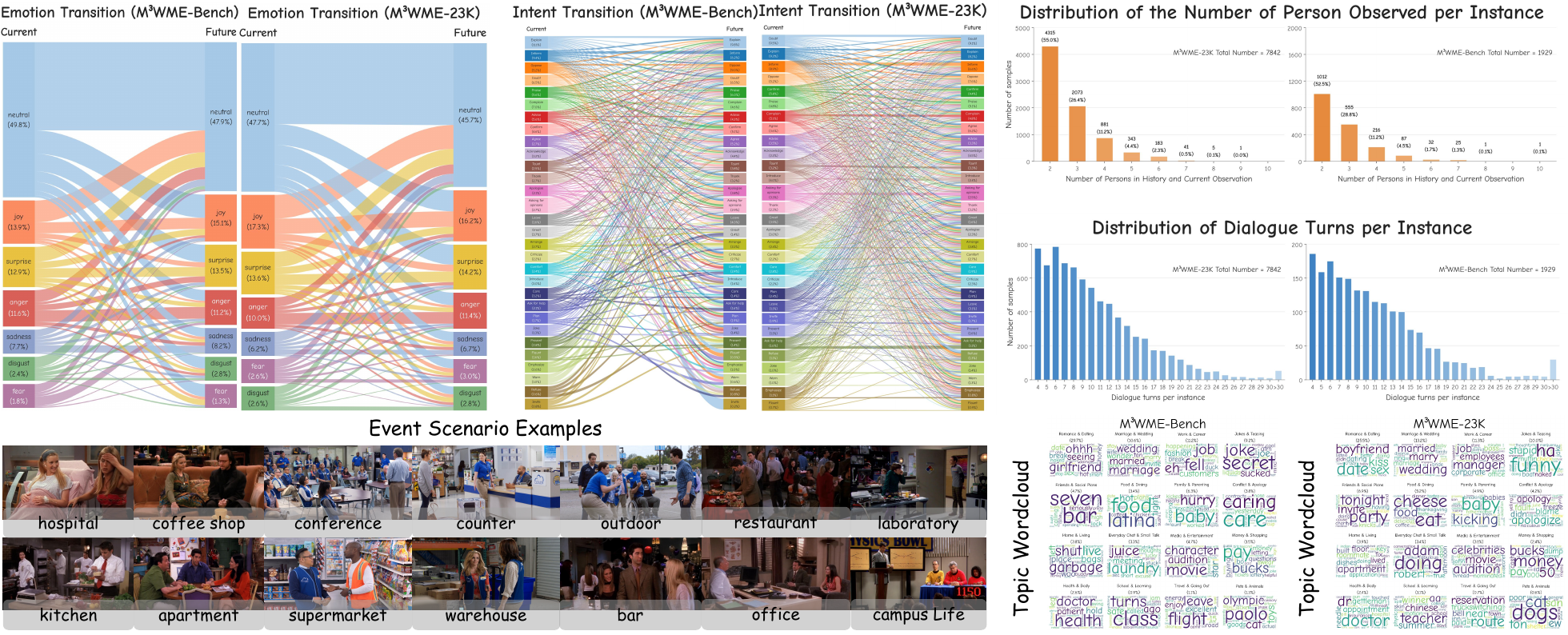}
    \caption{\textbf{Instance-level statistics of \benchname and M$^3$WME-23K.} \textbf{The upper-left part} illustrates the transitions of target subjects' mental states from current to future states. \textbf{The upper-right part} shows the distributions of participant counts and conversational turns across interaction instances. \textbf{The lower-left and lower-right parts} present common scenarios and conversation topics respectively.}
    \label{fig:data_distribution}
\end{figure*}
\subsection{Task Formulation}
\label{sec:task_formulation}
The primary objective of \benchname is to evaluate the ability of multimodal foundation models and agents to predict the future mental states and behaviors of a target subject, based on historical and current observations of multi-party interaction. This capability serves as the foundation for AI to integrate better into social environments and enable proactive human-robot interaction.

First, we factorize the world state $\mathcal{S}$ into observable physical states $\mathcal{P}$ and latent mental states $\mathcal{M}$ that cannot be directly observed but must be inferred from external cues, such that $\mathcal{S} = \mathcal{P} \times \mathcal{M}$. Our benchmark specifically focuses on emotion $\mathcal{E}$ and intention $\mathcal{I}$, defining the mental state space as $\mathcal{M} = \mathcal{E} \times \mathcal{I}$. 
Second, we denote the target subject as $\tau$ and the other interacting peers as $-\tau = \{i, j, \dots\}$. The task concentrates solely on the predictive mental and behavioral modeling of a single target individual. Regarding temporal granularity, we define the prediction target as the turn-level segment $t+1$ where the focal subject responds to others, rather than an instantaneous timestamp. The time window $t$ begins with the target subject's emergence and lasts until immediately prior to their next appearance.

During execution, the model accumulates historical memory of interaction $\mathcal{P}_{<t}$, observes the target's current behavior and verbal observation $\mathcal{P}_t^\tau$, and ingests subsequent peer observations $\mathcal{P}_t^{-\tau}$. The model is then required to predict the target's future mental state $\mathcal{M}_{t+1}^\tau$ and behavior $\mathcal{P}_{t+1}^\tau$. This process is formalized as follows:
\begin{equation}
\begin{aligned}
\begin{aligned}
\hat{\mathcal{M}}_{t+1}^\tau &= \arg\max_{\mathcal{M}_{t+1}^\tau} P_\theta\left(\mathcal{M}_{t+1}^\tau \mid \mathcal{P}_{\le t}, \tau\right), \\
\hat{\mathcal{P}}_{t+1}^\tau &= \arg\max_{\mathcal{P}_{t+1}^\tau} P_\theta\left(\mathcal{P}_{t+1}^\tau \mid \mathcal{P}_{\le t}, \tau\right).
\end{aligned}
\end{aligned}
    \label{eq:task_formulation}
\end{equation}

\subsection{Data Construction}

\noindent\textbf{Data Collection.} 
To systematically evaluate mental world modeling in multi-turn, multi-party contexts, we establish three essential criteria: (1) multi-modal interaction involving at least two active participants, (2) realistic everyday scenarios with rich topical diversity, and (3) dynamic mental states that evolve with conversational context. To satisfy these prerequisites, we leverage MELD~\cite{poria2019meld} and MIntRec2.0~\cite{zhang2024mintrec2} as our primary data foundations, which originate from three popular television series: \textit{Friends}, \textit{The Big Bang Theory}, and \textit{Superstore}. These corpora provide realistic everyday conversations equipped with speaker identity tags, dialogue transcriptions, and clip-level annotations for 7 emotion and 30 intention categories. The dense alternation between interlocutors and the strong contextual dependency make them ideal sources for our benchmark.

\noindent\textbf{Diversity Coverage.} 
Building upon these sources, we selectively sample conversational segments featuring alternating turns between the focal target subject and interacting peers. Crucially, we filter out continuous monologues and self-contained mental transitions where the target speaks uninterruptedly. This sampling strategy reinforces our fundamental assumption that the target's mental state transition is causally driven by others' preceding social actions. As depicted in \Cref{fig:data_distribution}, the resulting dataset spans diverse environments, \eg supermarkets, cafes, residential apartments, checkout counters, offices, university campuses, and research laboratories. Interpersonal relationships encompass friends, colleagues, and supervisor-subordinate dynamics, while conversational topics cover workplace interactions, customer service, romantic relationships, career development, scientific discussions, and geek culture. Spanning a spectrum from public workplaces to private domestic spaces, the dataset covers diverse everyday social scenarios involving multiple participants relevant to embodied agents and personal assistants.

\noindent\textbf{Data Curation Pipeline.} 
\begin{figure}[t]
    \centering
    \includegraphics[width=1.0\linewidth]{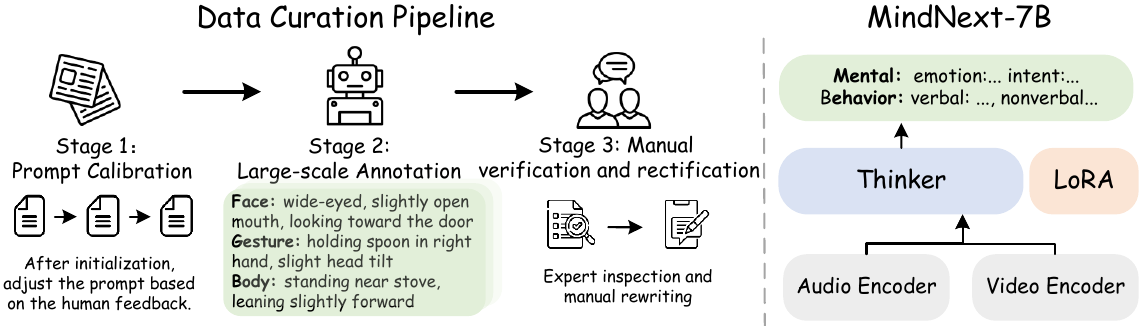}
    \caption{Overview of the three-stage data curation pipeline (left) and the omni-modal architecture of our MindNext-7B (right).}
    \label{fig:method}
\end{figure}
Since reliable emotion and intention annotations already exist in the raw corpora, our annotation pipeline concentrates on the target subject's prospective physical behaviors. As shown in \Cref{fig:method}, we deploy a three-stage workflow comprising prompt calibration, automated multimodal annotation, and manual verification:

\textit{Stage 1: Prompt Calibration.} We formulate an initial prompt and utilize Qwen3-VL-32B~\cite{bai2025qwen3vl} to execute pilot annotations on 50 representative samples. Human experts evaluate the descriptions generated  with respect to action precision, body orientations, facial expressions, and hallucination details, iteratively refining the prompt through three cycles to produce a standardized annotation protocol.

\textit{Stage 2: Large-Scale Annotation.} We employ Qwen3-VL-32B to generate full-scale behavioral descriptions, systematically capturing the focal subject's facial expressions, gestures, and body postures.

\textit{Stage 3: Manual Verification and Rectification.} To ensure annotation quality, annotators scrutinized the generated results. Our inspection primarily revealed three typical failure modes in model-generated annotations: misdirected actions toward incorrect interactants, action omissions, and hallucinated interactive objects. Subsequently, we devised rule-based heuristics to flag high-risk samples, manually rectifying 72.5\% of them, specifically addressing misdirected interactants and omitted actions. For validation, three expert annotators audited 100~stratified randomly sampled instances, yielding a 93\% pass rate (Krippendorff's $\alpha = 0.73$). This rigorous workflow provides a solid data base for subsequent training and evaluation.

Building upon this pipeline, we reserve the test split exclusively for benchmarking. Within this suite, we design targeted queries centered on future mental states and behavior forecasting, establishing \benchname. Meanwhile, the training and validation sets are augmented across diverse contextual configurations. Specifically, we introduce three complementary tasks: future mental modeling, trajectory-augmented mental prediction, and action forecasting, culminating in the final M$^3$WME-23K dataset.

\subsection{Data Statistics}

\Cref{fig:data_distribution} illustrates the transition distributions from present to future mental states in M$^3$WME-23K and \benchname, alongside the distributions of interactant counts and conversational turns per instance within the observed context. In general, the training and benchmark splits exhibit closely aligned data distributions. As summarized in \Cref{tab:statistics}, the M$^3$WME-23K training suite comprises 23,532 QA samples spanning approximately 169 hours of video, 201K conversational turns, and 228 distinct target personas. In comparison, \benchname contains 3,858 QA instances and 34K conversational turns covering 14.29 hours of video. Together, these comprehensive resources establish a solid empirical foundation for both rigorous benchmarking and effective baseline training.

\begin{table}[t]
    \centering
    \caption{Statistics of \benchname and $\text{M}^{3}$WME-23K.}
    \label{tab:statistics}
    \footnotesize
    \resizebox{\linewidth}{!}{%
    \begin{tabular}{lcc}
        \toprule
        Statistics & \benchname & $\text{M}^{3}$WME-23K \\
        \midrule
        Data Usage     & Test           & Train \& Val  \\
        Video          & 1,929          & 7,844       \\
        QA Samples     & 3,858          & 23,532      \\
        Target Subject & 99             & 228         \\
        Utterance      & 34K            & 201K        \\
        Duration       & 14.29 h        & 169.04 h \\
        \bottomrule
    \end{tabular}
    }
\end{table}

\subsection{Evaluation Metrics and Protocol}
\noindent{\textbf{The PMM Metrics.}} To account for category imbalance across state transitions, we evaluate predictive mental modeling using Weighted Average F1 (WAF), Weighted Precision (WP), and Weighted Recall (WR) across both emotion and intention dimensions.

\noindent{\textbf{The PBM Metrics.}} We evaluate both verbal and non-verbal behavioral prediction using two primary metrics as follows:

\textit{Behavioral Semantic Similarity (BSS).} We employ the sentence representation model all-mpnet-base-v2~\cite{song2020mpnet} to compute cosine semantic similarities across four distinct behavioral components: spoken utterance, facial expression, hand gesture, and body posture.
\textit{LLM-as-a-Judge (LLM-Judge).} We deploy CompassJudger2-32B~\cite{zhang2025compassjudger2} governed by rigorous rubric prompts. The judge model independently scores each of the four components on a scale of 0 to 100 divided into six discrete performance levels: 0, 20, 40, 60, 80, and 100. See the appendix for more details.

Both BSS and LLM-Judge metrics aggregate the verbal and non-verbal sub-scores into an overarching action score using a unified formulation:
\begin{equation}
    \text{Overall} = 0.5 \times s_{\text{verbal}} + 0.5 \times \text{mean}\left(s_{\text{face}}, s_{\text{gesture}}, s_{\text{body}}\right),
    \label{eq:score_agg}
\end{equation}
where $s_{\text{verbal}}$ denotes the utterance prediction score, and the non-verbal score is calculated as the arithmetic mean across facial $s_{\text{face}}$, gestural $s_{\text{gesture}}$, and postural $s_{\text{body}}$ dimensions.

\noindent{\textbf{Protocol.}} 
To ensure a rigorous and fair benchmark, we institute an adaptive and fair evaluation protocol that structures the input context into historical observations, current observations, and task instructions. Specifically, historical observations are provided as textual dialogue memories of the characters, while the integration of multimodal historical contexts is systematically investigated in \Cref{sec:further_analysis}. For end-to-end models, we evaluate under the \textit{Pass@1} setting with the sampling temperature set to 0, examining the success rate of a single deterministic prediction attempt. For agentic frameworks, we employ a unified backbone to ensure a fair comparison across different systems. Furthermore, agentic methods are permitted to conduct multi-turn intermediate artifact generation, tool invocations, and LLM-based verification before producing predictions. To align with our benchmark, we retain the core mechanisms of representative Theory-of-Mind (ToM) frameworks while adapting their downstream output interfaces. To ensure standardized and automated evaluation, all models are strictly required to respond following a predefined structured format. See the Appendix for more details.

\subsection{Baselines}
As shown in \Cref{fig:method}, capitalizing on the proposed M$^3$WME-23K, we perform supervised fine-tuning on the omni-modal language backbone Qwen2.5-Omni-7B to establish MindNext-7B as a strong baseline. MindNext-7B natively ingests multi-turn conversational contexts consisting of interleaved video, audio, and textual streams across multiple participants, forecasting the target subject's future mental state and behaviors via task-specific queries. 

To establish a holistic benchmark landscape, we evaluate four representative paradigms: general MLLMs that excel at broad multimodal reasoning, emotion-specialized MLLMs optimized for emotion recognition and cause analysis, agentic frameworks encompassing tool-augmented video agents and ToM-based cognitive deliberation systems, and world foundation models spanning multimodal video generation and embodied comprehension. We also include a State Persistence Baseline that directly copies the ground-truth current mental state as the future prediction.

%% file: sec/4_experiment.tex
\section{Experiments}
\begin{table*}[t]
  \centering
  \caption{\textbf{Comparison with mainstream approaches on predictive mental and behavioral modeling.} The best result is highlighted in \textbf{bold}, and the best result within each category is marked with \colorbox{cvprblue!25}{shading} . $\dagger$ denotes retention of the core mechanism, adapted for task execution. Avg. denotes the arithmetic mean calculated for the Emotion \& Intention WAF, Overall BSS \& LLM-Judge metrics.}
  \label{tab:main_results}
  \renewcommand{\arraystretch}{1}
  \setlength{\tabcolsep}{2.1pt}
  \footnotesize
  \begin{tabular}{@{}l c|MMM|MMM|cccccc|c@{}}
    \toprule
    \multirow{3}{*}{Methods}
    & \multirow{3}{*}{Modality}
    & \multicolumn{6}{c|}{Predictive Mental Modeling}
    & \multicolumn{6}{c|}{Predictive Behavioral Modeling}
    & \multirow{3}{*}{Avg.} \\
    \cmidrule(lr){3-14}
    & & \multicolumn{3}{c}{Emotion}
    & \multicolumn{3}{c|}{Intention}& \multicolumn{2}{c}{Verbal} & \multicolumn{2}{c}{Non-verbal}& \multicolumn{2}{c|}{Overall}  &\\
    \cmidrule(lr){3-5} \cmidrule(lr){6-8} \cmidrule(lr){9-10} \cmidrule(lr){11-12} \cmidrule(lr){13-14}
    & & WAF & WP & WR & WAF & WP & WR & BSS & LLM-Judge & BSS & LLM-Judge & BSS & LLM-Judge & \\
    \midrule
    \multicolumn{10}{@{}l}{\textit{General MLLMs}} \\
    \midrule
    GLM-4.6V-Flash~\cite{hong2025glm45v}  &VT & 34.10 & 43.07 & 31.95 & 6.61 & 9.32  & 6.83 &18.74 &10.19 &\cellcolor{cvprblue!25}66.93 &\cellcolor{cvprblue!25}41.35 & 42.83 &\cellcolor{cvprblue!25}25.77 &\cellcolor{cvprblue!25}27.33 \\
    Qwen3.5-9B~\cite{qwen3.5}      &VT & 32.01 & 39.75 & 31.25 & \cellcolor{cvprblue!25}6.83 & 9.01  & \cellcolor{cvprblue!25}9.52 &20.25 &10.20 &65.71 &36.44  & \cellcolor{cvprblue!25}42.98 &23.33 &26.29 \\
    Qwen3-Omni-30B~\cite{xu2025qwen3omni}  &AVT & 31.02 & \cellcolor{cvprblue!25}44.15 & 30.79 & 6.80 & \cellcolor{cvprblue!25}11.19  & 8.10 &\cellcolor{cvprblue!25}\textbf{22.74} &\cellcolor{cvprblue!25}11.64 &62.02 &32.74   & 42.38 &22.19 &25.60 \\
    Qwen2.5-Omni-7B~\cite{xu2025qwen25omni} &AVT & \cellcolor{cvprblue!25}36.57 & 42.43 & \cellcolor{cvprblue!25}37.03 & 3.19 & 10.07 & 6.03 &20.27 &8.18 &59.55 &29.90  & 39.91 &19.04 &24.68 \\
    Kimi-VL-A3B~\cite{du2025kimivl}     &VT & 34.12 & 41.47 & 33.87 & 6.68 & 10.15 & 7.46 &15.54 &6.81 &42.21 &28.78  & 28.87 & 17.79 & 21.87 \\
    \midrule
    \multicolumn{10}{@{}l}{\textit{Emotion MLLMs}} \\
    \midrule
    OneEmo~\cite{huang2026oneemo}          &VT & 35.08 & 40.01 & 34.03 & \cellcolor{cvprblue!25}6.59 & 6.12 & \cellcolor{cvprblue!25}9.37 &17.11 &\cellcolor{cvprblue!25}7.04 &\cellcolor{cvprblue!25}53.68 &\cellcolor{cvprblue!25}29.37  & 35.39 &\cellcolor{cvprblue!25}18.26 &\cellcolor{cvprblue!25}23.83 \\
    EmotionThinker~\cite{wang2026emotionthinker}  &AVT & \cellcolor{cvprblue!25}37.94 & 36.49 & \cellcolor{cvprblue!25}41.34 & 3.90 & 6.05  & 5.71 &\cellcolor{cvprblue!25}20.79 &6.05 &50.90 &9.11  & \cellcolor{cvprblue!25}35.84 &7.57 &21.31 \\
    VidEmo-7B~\cite{zhang2025videmo}       &VT & 18.36 & 49.38 & 14.32 & 2.22 & 4.42  & 3.33 &13.89 &3.43 &37.53 &13.74  & 25.70 &8.57 &13.71 \\
    AffectGPT~\cite{lian2025affectgpt}       &AVT & 16.75 & 39.79 & 14.86 & 2.45 & \cellcolor{cvprblue!25}16.74 & 2.38 &7.68 &2.87 &20.21 &9.59  & 13.94 &6.23 &9.84 \\
    SABER-LLM~\cite{zhao2026saber}       &AVT & 8.05  & \cellcolor{cvprblue!25}\textbf{49.73} & 14.86 & 1.37 & 1.06  & 3.97 &11.45 &2.06 &16.33 &3.63  & 13.88 &2.84 &6.54 \\
    \midrule
    \multicolumn{10}{@{}l}{\textit{Agentic Frameworks~(Qwen3.8-27B~\cite{qwen2026qwen38})}} \\
    \midrule
    VideoTree~\cite{yu2025videotree}       &VT & \cellcolor{cvprblue!25}40.19 & \cellcolor{cvprblue!25}44.36 & \cellcolor{cvprblue!25}38.88 & \cellcolor{cvprblue!25}8.86 & 10.50 & \cellcolor{cvprblue!25}9.68 &20.87 &10.38 &64.54 &33.06  & \cellcolor{cvprblue!25}42.70 &21.76 &\cellcolor{cvprblue!25}28.38 \\
    VideoARM~\cite{yin2026videoarm}        &AVT & 34.17 & 38.37 & 32.49 & 8.16 & 9.08  & 8.73 &\cellcolor{cvprblue!25}21.69 &\cellcolor{cvprblue!25}11.81 &61.99 &\cellcolor{cvprblue!25}34.14  & 41.84 &\cellcolor{cvprblue!25}22.99 &26.79 \\
    MetaMind$^{\dagger}$~\cite{zhang2025metamind} &T & 35.37 & 38.09 & 34.33 & 6.48 & 7.11  & 8.10 &20.61 &9.96 &63.11 &31.57  & 41.86 &20.79 &26.13 \\
    LIMP$^{\dagger}$~\cite{shi2025muma}     &T & 32.10 & 39.28 & 29.41 & 5.08 & 6.41  & 6.03 &20.55 &9.87 &62.68 &31.15  & 41.61 &20.53 &24.83 \\
    Mentis$^{\dagger}$~\cite{fei2026mental}   &VT & 30.47 & 37.66 & 29.18 & 5.07 & \cellcolor{cvprblue!25}11.84 & 6.67 &20.10 &7.03 &\cellcolor{cvprblue!25}65.29 &34.01  & 42.69 &20.53 &24.69 \\
    \midrule
    \multicolumn{10}{@{}l}{\textit{World Foundation Models}} \\
    \midrule
    RynnBrain1.1-9B~\cite{li2026rynnbrain} &VT & \cellcolor{cvprblue!25}35.47 & 40.90 & \cellcolor{cvprblue!25}33.64 & \cellcolor{cvprblue!25}9.04 & 9.99  & \cellcolor{cvprblue!25}10.16 &\cellcolor{cvprblue!25}20.41 &\cellcolor{cvprblue!25}8.78 &64.77 &34.68 & \cellcolor{cvprblue!25}42.59 &21.76 &\cellcolor{cvprblue!25}27.22 \\
    RoboBrain2.5-8B~\cite{tan2026robobrain}  &VT  & 26.40 & 39.80 & 27.79 & 7.06 & 7.05  & 9.68 &20.36 &8.01 &62.54 &34.46  & 41.45 &21.26 &24.04 \\
    Cosmos3-Nano~\cite{nvidia2025cosmos}    &VT & 23.57 & 40.95 & 25.56 & 6.41 & 7.66  & 7.94 &20.09 &7.73 &64.35 &33.26  & 42.21 &20.53 &23.18 \\
    Emu3.5~\cite{cui2025emu35}         &VT & 14.10 & 36.44 & 20.25 & 5.50 & \cellcolor{cvprblue!25}\textbf{18.64} & 7.14 &18.10 &8.12 &\cellcolor{cvprblue!25}65.80 &\cellcolor{cvprblue!25}36.37  & 41.95 &\cellcolor{cvprblue!25}22.27 &20.96 \\
    BAGEL-7B-MoT~\cite{deng2025bagel}    &VT & 11.89 & \cellcolor{cvprblue!25}49.20 & 18.40 & 4.80 & 8.69  & 6.03 &17.68 &6.90 &53.30 &32.79  & 35.49 &19.85 &18.01 \\
    \midrule
    State Persistence Baseline &- &40.87 &40.61 &41.19 &9.87 &10.32 &10.16 &- &- &- &- &- &-  &-\\
    \textbf{Ours MindNext-7B}      &AVT &\textbf{42.99} &44.39 &\textbf{49.27} &\textbf{11.83} &10.35 &\textbf{15.24} &22.42 &\textbf{13.29} &\textbf{68.04} &\textbf{43.90}  &\textbf{45.22} &\textbf{28.60} &\textbf{32.16} \\
    \bottomrule
  \end{tabular}%
\end{table*}
\subsection{Main Results}

In this section, we present an extensive comparative evaluation across the four representative paradigms. \Cref{tab:main_results} reports the comprehensive performance across all evaluation dimensions and task formulations.

\noindent{\textbf{General MLLMs.}} 
Among general vision-language architectures, GLM-4.6V-Flash delivers the best overall performance, demonstrating a balanced capability in forecasting both future mental states and prospective behaviors. In contrast, omni-modal models fall behind pure vision-language baselines on non-verbal behavioral prediction and overall score, despite Qwen3-Omni-30B leading in verbal forecasting. Thus, across current architectures, neither incorporating audio streams nor scaling omni-modal capacity automatically translates into stronger holistic behavioral foresight. Overall, open-source MLLMs still encounter significant performance bottlenecks in forecasting future psychological transitions and prospective behavior.

\noindent{\textbf{Emotion MLLMs.}} 
Within the affective computing MLLM paradigm, the 4.5B OneEmo achieves the highest average score, followed closely by the 7B omni-modal EmotionThinker. Crucially, these specialized models, despite being state-of-the-art on conventional emotion analysis, recognition and causal reasoning benchmarks, fail to outperform general MLLMs in predictive mental-behavioral modeling. We attribute this discrepancy to an alignment shift induced by static affective post-training, which over-specializes models for instantaneous cue fitting at the expense of forward simulation.

\noindent{\textbf{Agentic Frameworks.}} 
Agentic methods demonstrate competitive results, with VideoTree attaining the highest performance by leveraging tree-based frame selection. Among Theory-of-Mind (ToM) systems, MetaMind, which formulates hypotheses and psychological constraints over interactants, yields notable competence. Nonetheless, architectures that explicitly instantiate mental states or infer mental states inversely from observed behaviors fail to surpass visual frame selection agents. This performance discrepancy stems from the fact that existing ToM agents predominantly synthesize current mental observations rather than generating informative increments for future forecasting. Although video agents effectively filter salient visual clips via tool calling and frame preprocessing, they produce marginal improvements in predictive mental and behavioral modeling.

\noindent{\textbf{World Foundation Models.}} 
The embodied reasoning model RynnBrain1.1-9B achieves the best scores across both mental state and action prediction within this category, consistently outperforming Cosmos3-Nano and the 34B Emu3.5, thus demonstrating parameter efficiency and architectural advantages. Nevertheless, these models still trail behind our specialized MLLM baseline MindNext-7B and fail to exceed the general vision model GLM-4.6V-Flash. This performance lag exposes a fundamental gap between physical world modeling and mental world modeling. Architectures adept at multimodal generation, physical simulation, and spatial trajectory planning remain ill-equipped for human future mental state and behavior forecasting in open social environments.

Furthermore, we observed that all four paradigms perform notably worse in predicting future verbal behaviors than non-verbal behaviors, indicating that accurately forecasting the target agent's forthcoming dialogue content in multi-party, multi-turn interaction settings remains a significant challenge.

\subsection{Further Analysis}
\label{sec:further_analysis}
We conduct further behavioral and mechanistic analyses to identify the critical factors governing model performance, focusing on information sources, internal representations, and attribution dynamics.

\begin{figure}[t]
    \centering
    \includegraphics[width=1.0\linewidth]{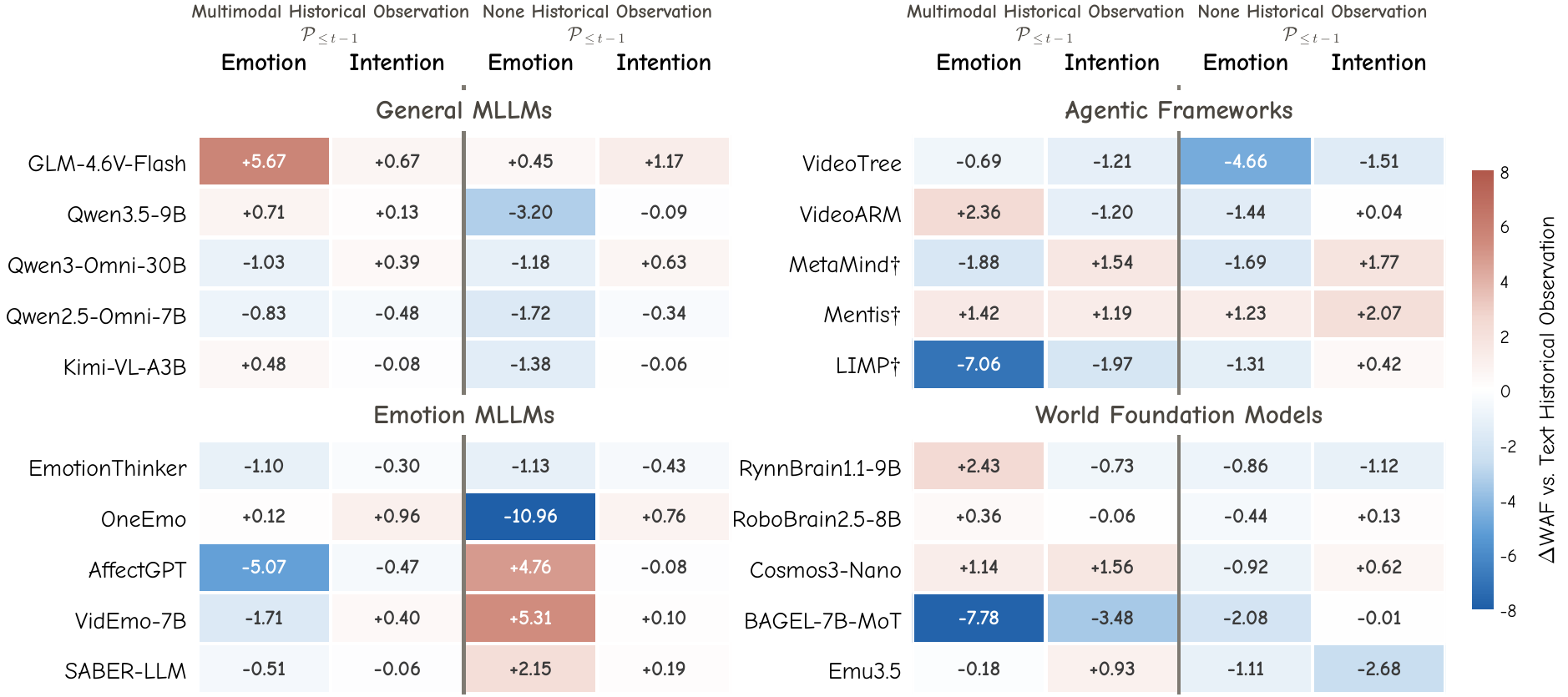}
    \caption{\textbf{Performance discrepancies across different historical observation settings.} Results are reported relative to the text-only memory baseline, where \textcolor{red}{Red} and \textcolor{blue}{Blue} indicate performance degradation and improvement relative to the baseline, respectively.}
    \label{fig:history_analysis}
\end{figure}
\noindent \textbf{Historical Memory $\mathcal{P}_{\leq{t-1}}$ Remains Under-utilized.} 
To investigate how history context affects model predictions of prospective mental states, we examine two configurations: augmenting history with full audio-visual modalities and removing historical observations altogether. As depicted in \Cref{fig:history_analysis}, incorporating multimodal historical contexts yields modest improvements for most general MLLMs. Notably, GLM-4.6V-Flash achieves substantial gains in prospective emotion forecasting, whereas most emotion MLLMs experience performance drops. Agentic frameworks and world foundation models exhibit uneven fluctuations, where LIMP and BAGEL-7B-MoT decline markedly in mental forecasting. This pattern suggests that multimodal observations can introduce an additional informational burden, preventing both end-to-end and agentic methods from consistently extracting actionable cues. 

In contrast, completely ablating historical memory causes general MLLMs to degrade in emotion forecasting while showing marginal gains in intent prediction. Emotion MLLMs diverge sharply upon history removal: while AffectGPT, VidEmo-7B, and SABER-LLM improve markedly on emotion, EmotionThinker and especially OneEmo degrade. Agentic frameworks and world foundation models similarly exhibit mild perturbations. These divergent behaviors likely reflect the post-training objectives of emotion MLLMs, which heavily prioritize static mental perception, meaning extended historical contexts may dilute their signal-to-noise ratio. Meanwhile, other paradigms tend to exploit historical contexts only selectively. In summary, these findings demonstrate that current approaches fail to reliably and fully mine the informative signals embedded within historical memory.

\begin{figure}[t]
    \centering
    \includegraphics[width=1.0\linewidth]{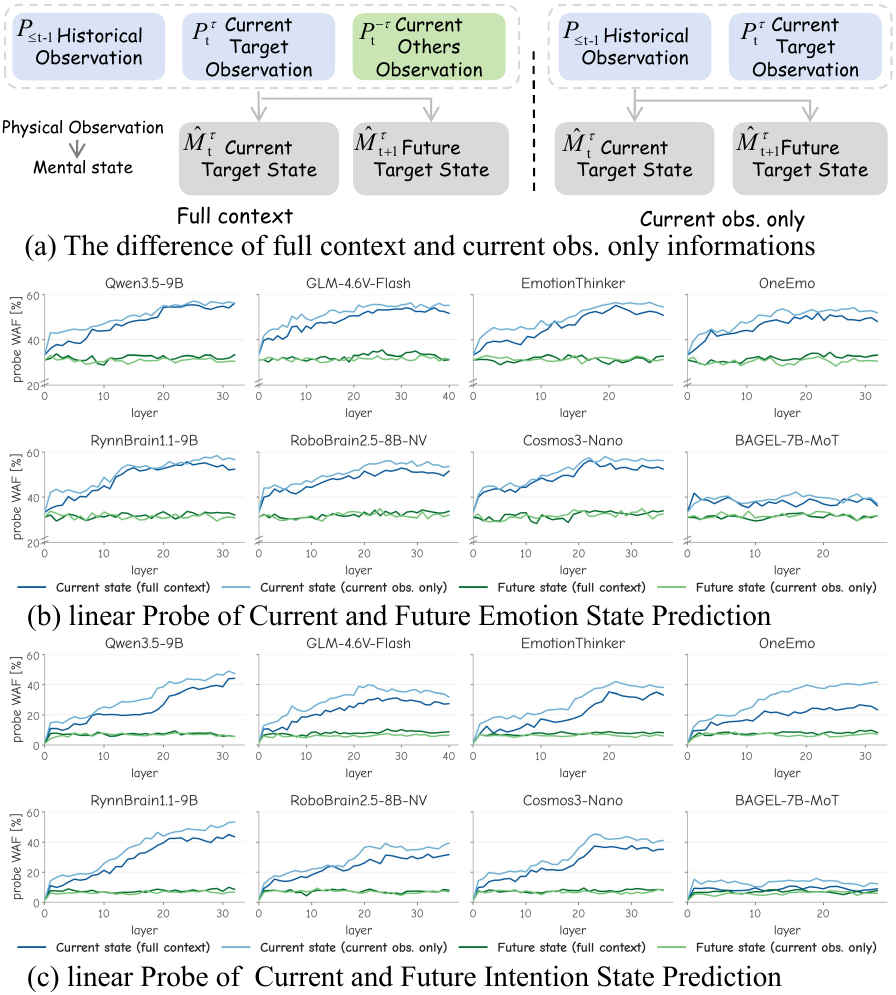}
    \caption{\textbf{Probing mental representations under different context utilization.} (a) Explains the differences in various contextual information. (b, c) Layer-wise linear probing reveals that existing methods consistently fail to encode future mental state.}
    \label{fig:linear_probe}
\end{figure}

\begin{figure*}[t]
    \centering
    \includegraphics[width=1.0\linewidth]{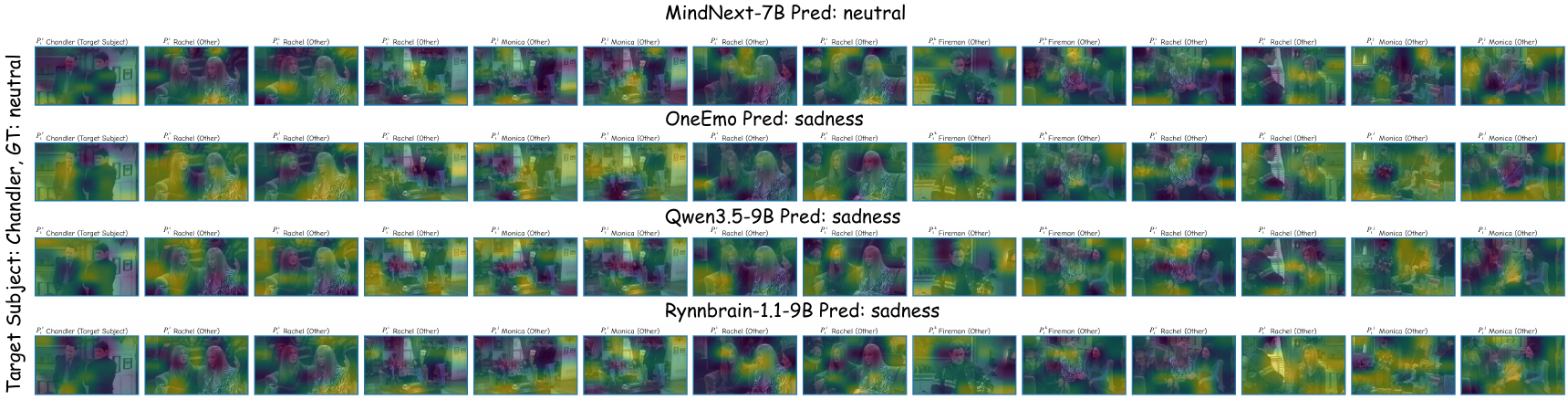}
    \caption{\textbf{Visual attribution analysis on current observation.}  Brighter regions indicate higher model reliance on the corresponding visual cues, whereas darker areas reflect lower contributions toward forecasting prospective emotion.}
    \label{fig:atttribution_analysis}
\end{figure*}

\begin{figure}[ht]
    \centering
    \includegraphics[width=1.0\linewidth]{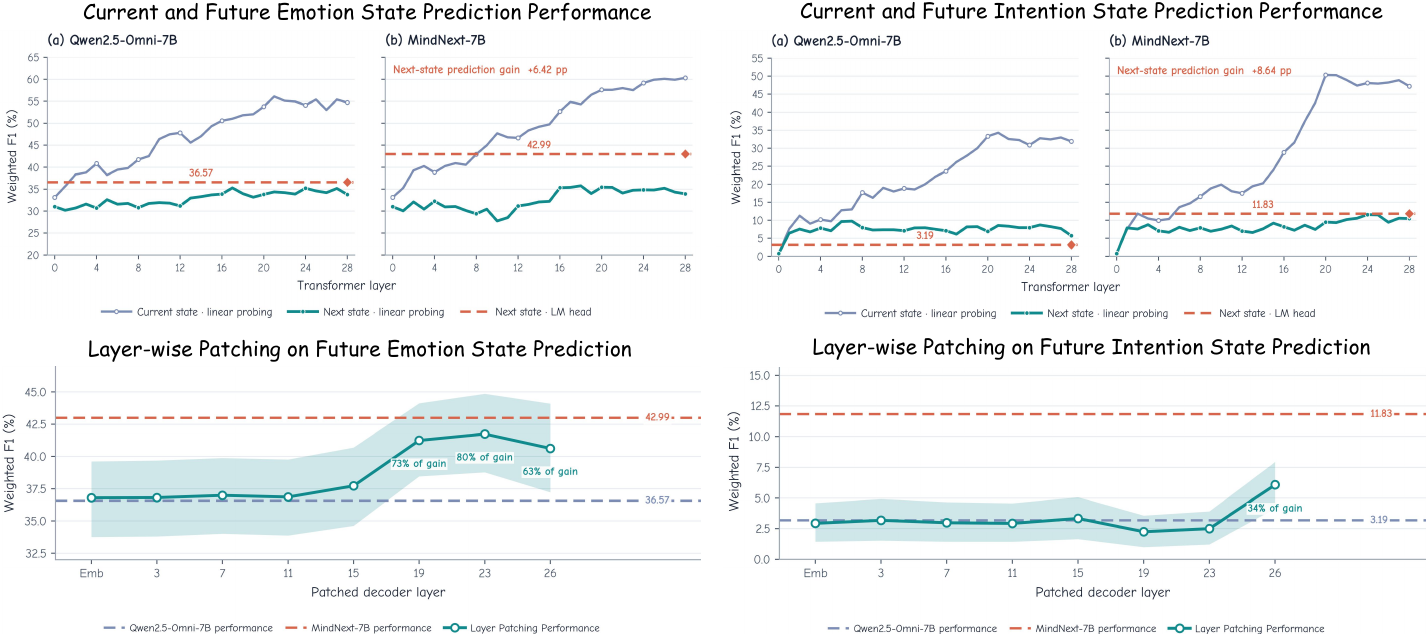}
    \caption{\textbf{The upper panels} show the linear probing and language model (LM) head readout performance for mental state prediction. \textbf{The lower panels} show the results of layer-wise representation patching across the backbone.}
    \label{fig:linear_probe_sft}
\end{figure}
\noindent\textbf{A primary performance bottleneck lies in the failure to transform observations into decision-ready future representations.} 
This bottleneck stems from two possible sources. It reflects either insufficient future state computation, or a failure of the output head to exploit existing latent information.
We conducted a linear probing analysis on the residual stream of each layer of the model.
As depicted in \Cref{fig:linear_probe}, a stark representational asymmetry emerges: \textit{whereas current mental states become increasingly decodable across intermediate and deep layers in every architecture, future state decodability remains persistently depressed and largely invariant across layer depth.} 

Furthermore, when ablating subsequent behavioral observations of interacting peers, the accuracy of current mental inference improves noticeably, which can be attributed to the elimination of extraneous contextual distractors. In contrast, this ablation introduces only marginal perturbations to future mental forecasting.

These empirical results reveal key limitations in existing architectures. Specifically, they fail to explicitly model future mental state. Moreover, they struggle to condition future predictions on others' social actions.

\textbf{Supervised Fine-Tuning (SFT) Converts Weakly Accessible Latent Signals into Output-ready Representations.} \Cref{fig:linear_probe_sft} compares external linear probing with LM head readout before (Qwen2.5-Omni-7B) and after SFT (MindNext-7B). We find that the model's internal representations of future mental states exhibit negligible improvement under external linear probing after fine-tuning; in contrast, readout head performance via the fine-tuned language model improves substantially. We patch MindNext-7B's hidden representations layer by layer into Qwen2.5-Omni-7B and find that patching into deeper layers partially restores performance. These results indicate that SFT does not substantially enhance the external linear decodability of future states. Instead, it reshapes hidden representations at decision points, aligning latent future-relevant cues with the fixed readout directions of the LM head.

\noindent\textbf{Accurate forecasting requires identifying critical cues amid complex social contexts.} 
To better understand the role of current visual observations, we perform a model-agnostic attribution analysis that requires neither gradients nor attention maps~\cite{chen2026mllms}. \Cref{fig:atttribution_analysis} shows the attribution of visual inputs to the final prediction. MindNext-7B sharply focuses on critical cues, specifically character facial expressions and gestures, whereas baseline methods are frequently distracted by irrelevant background. As a result, our model successfully infers Chandler's future emotional state. In future work, better exploiting observed multi-turn social interactions remains a promising avenue for improving mental world modeling.
For more text and visual attribution visualization analysis, please refer to the Appendix.

%% file: sec/5_conclusion.tex
\section{Conclusion}
In this work, we introduce \benchname, a novel benchmark dedicated to evaluating explicit modeling of human mental state transitions and behavioral forecasting. Distinct from conventional static mental perception tasks, our benchmark focuses on dynamic, multi-party, and multi-turn interactions, requiring models to prospectively forecast a target subject's future mental states as well as verbal and nonverbal behavior. The benchmark comprises 99 distinct target personas and 3,858 structured QA pairs, supported by a standardized evaluation protocol. In addition, we construct the M$^3$WME-23K training suite spanning 228 personas, 23,532 QA pairs, and establish MindNext-7B as a competitive omni-modal LLM baseline for future research. Extensive evaluations across four representative paradigms demonstrate that contemporary architectures struggle to forecast future mental states and behaviors, despite their mastery of multimodal comprehension, affective understanding, video generation, and physical planning. We envision \benchname serving as a foundational human-centric testbed to catalyze the development of socially adaptive artificial intelligence. 

\section*{LLM Usage Statement}
We used Gemini-3.8-Flash for language refinement and GPT-5.6-Sol as an assistive tool for coding , implementation and data accuracy verification. All conceptualization, experimental design, and interpretation of the results were conducted exclusively by the authors.

%% file: sec/6_suppl.tex
\clearpage
\appendix
\setcounter{page}{1}
\setcounter{table}{0}
\setcounter{figure}{0}
\setcounter{section}{0}
\setcounter{equation}{0}
\renewcommand{\thesection}{\Alph{section}}

\renewcommand{\theHsection}{appendix.\thesection} 

\maketitlesupplementary

\section{Overview}
\label{supsec:overview}
As part of the supplementary material, we present the following extensions to the main paper:

\begin{itemize}[leftmargin=*, label=$\bullet$]
    \item Application Scenarios (\cref{supsec:application})
    \item Additional Details (\cref{supsec:additional-setup})
    \item Additional Experiments Analysis (\cref{supsec:additional-experiments})
    \item Prompt Protocol (\cref{supsec:prompts})
\end{itemize}

\section{Application Scenarios}
\label{supsec:application}
\begin{figure}[h]
  \centering
  \includegraphics[width=1.0\linewidth]{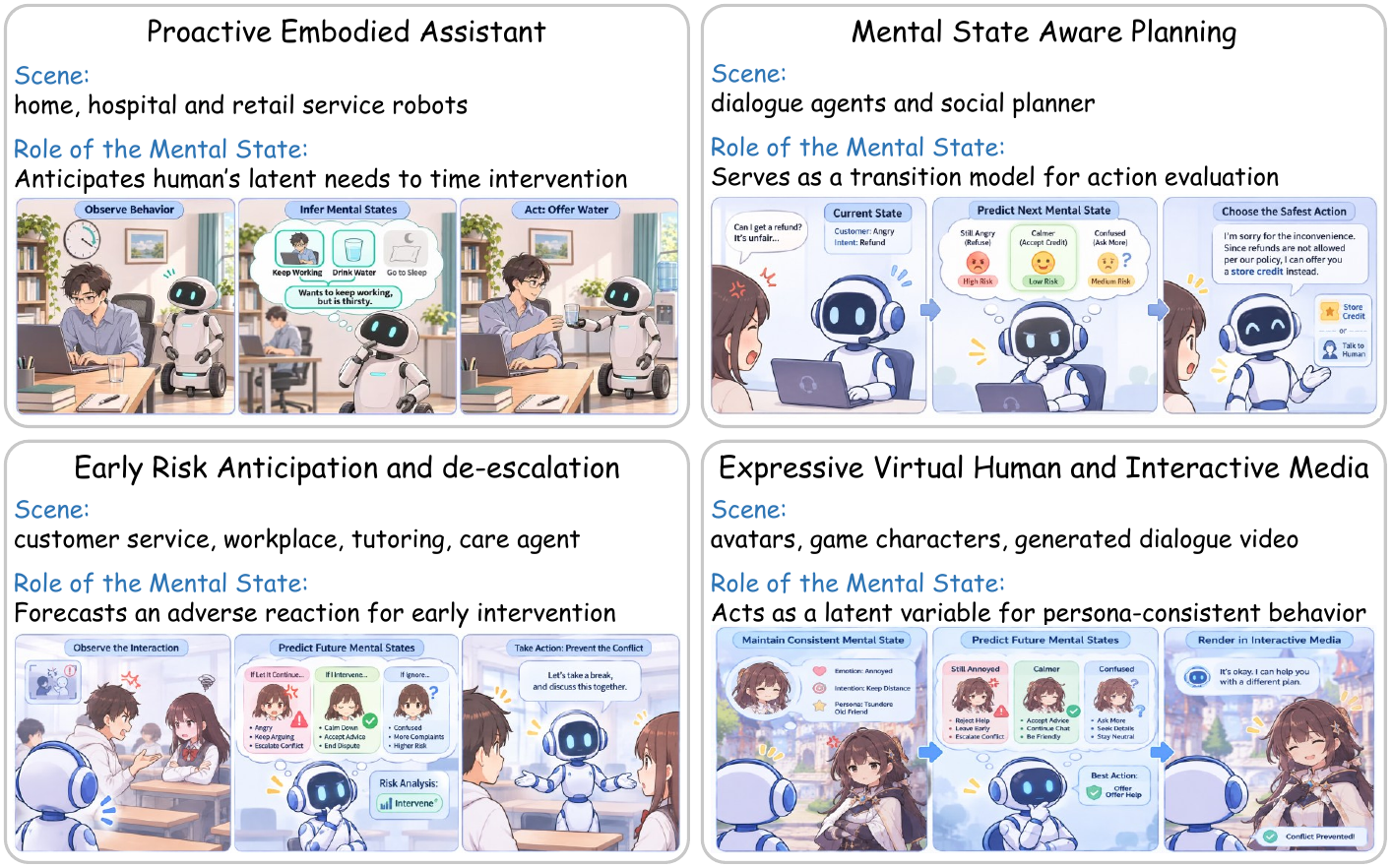}
  \caption{Candidate applications of mental modeling, including embodied action, mental-state-aware planning, risk prediction and intervention, and interactive virtual character generation.}
  \label{fig:app}
\end{figure}

\Cref{fig:app} illustrates four potential applications of multi-turn future mental state modeling. 

\noindent{\textbf{Proactive Embodied Assistant.}}
After observing a user working at a desk for an extended period, embodied AI predicts their future beliefs and intentions. Specifically, anticipating that the user is likely thirsty yet wants to keep working, the robot proactively fetches them a glass of water. 

\noindent{\textbf{Mental State Aware Planing.}}
To mitigate a user's poor shopping experience, a customer-service agent predicts the future emotional states each action may trigger and offers store credit instead of probing further.

\noindent{\textbf{Early Risk Anticipation and de-escalation.}}
When observing group interactions, if the robot assesses that an exchange is likely to escalate into conflict, it can proactively intervene to reduce the likelihood of subsequent conflictual behavior. 

\noindent{\textbf{Expressive Virtual Human and Interactive Media.}}
In generating game NPCs or interactive avatars, latent mental states can serve as an underlying factor to maintain consistent personas and natural emotions.

\section{Additional Details}
\label{supsec:additional-setup}

\noindent{\textbf{Linear Probe Setup.}} 
We empirically examine whether the core performance bottleneck stems from representational deficiencies inside the models, specifically testing our hypothesis regarding the lack of explicit mental state transition modeling. We select top-performing representatives across general MLLMs, emotion-specialized MLLMs, and world foundation models. Specifically, we extract layer-wise residual streams on M$^3$WME-23K and train linear probes to decode mental states. In \benchname, we instruct models to infer current mental states and to forecast prospective mental transitions, respectively.

\noindent{\textbf{Layer-wise Patching Setup.}}
We first cache the forward residual streams of Qwen2.5-Omni-7B and its fine-tuned counterpart, MindNext-7B, on M$^3$WME-Bench. Next, we apply single-layer patching at the final token position. For each target layer, we replace Qwen2.5-Omni-7B's residual-stream representation at the last token with that of MindNext-7B, while leaving all other layers unpatched. The token embeddings, attention, and MLP weights all remain fixed to Qwen2.5-Omni-7B. Through this intervention, we investigate at which layers fine-tuning gains reside, which layers enable effective readout, and how much performance can be restored.

\noindent{\textbf{Human Evaluation Setup.}}
To ensure a reliable evaluation of future behavior modeling, we conducted human validation in addition to automatic metrics. Specifically, we selected 100 instances via stratified random sampling to ensure sample balance. Two human evaluators scored the predicted verbal and nonverbal behaviors on a scale of 0 to 100 following the LLM-Score rubrics, with the overall score computed according to Eq. (2).

\noindent{\textbf{Visual Attribution Analysis Setup.}}
We employ a gradient-free interpretability method. Specifically, for each sample, the model first generates its own prediction. We then greedily insert (measuring sufficiency) and delete (measuring necessity) visual regions, mapping the model's predictive confidence back to the input regions. In practice, each frame is partitioned into a $5 \times 5$ grid, and the resulting regions are ranked according to their order of selection.

\section{Additional Experiments Analysis}
\label{supsec:additional-experiments}

\begin{figure}[ht]
    \centering
    \includegraphics[width=1.0\linewidth]{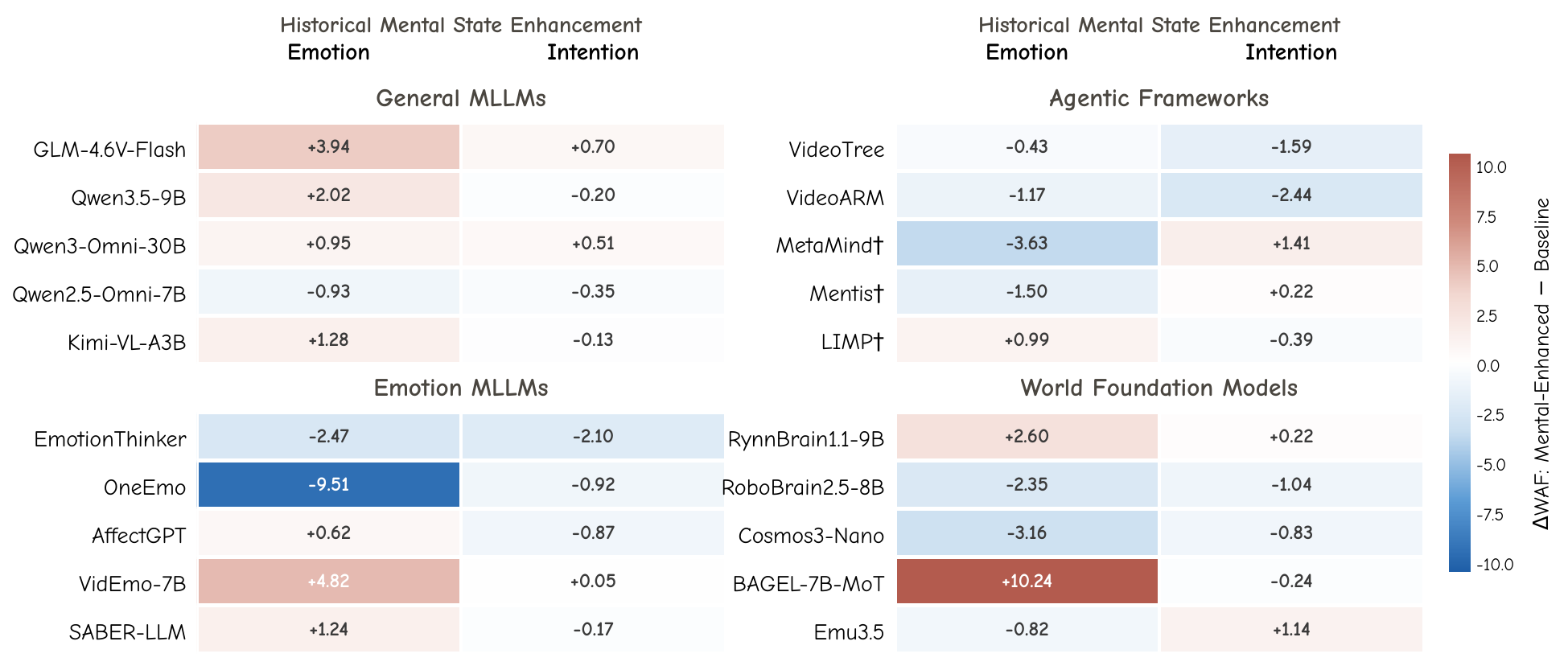}
    \caption{\textbf{Performance comparison} between the augmented setting (injecting historical mental trajectories of the target subject) and the baseline (pure textual dialogue memory).}
    \label{fig:historical-mental-enhanced}
\end{figure}

\noindent\textbf{Simply Reading Historical Mental Trajectories Falls Short.} 
Augmenting historical memory with the target agent's historical mental trajectories does not yield uniform improvements across all methods (\Cref{fig:historical-mental-enhanced}). While general MLLMs generally achieve positive net gains, emotion-oriented MLLMs exhibit mixed results, with EmotionThinker and OneEmo even degrading. Agent frameworks suffer net performance drops, yet the world foundation model BAGEL-7B-MoT shows shows substantial improvement on emotion prediction, presumably owing to superior reasoning capabilities from its pre-training. Overall, passive observation of historical mental trajectories does not guarantee better future mental state predictions across paradigms; rather, it introduces cognitive overhead for some models. This demonstrates that naive memory concatenation is insufficient, underscoring the necessity to explicitly process both physical and mental historical observations.
\begin{figure}[t]
    \centering
    \includegraphics[width=1.0\linewidth]{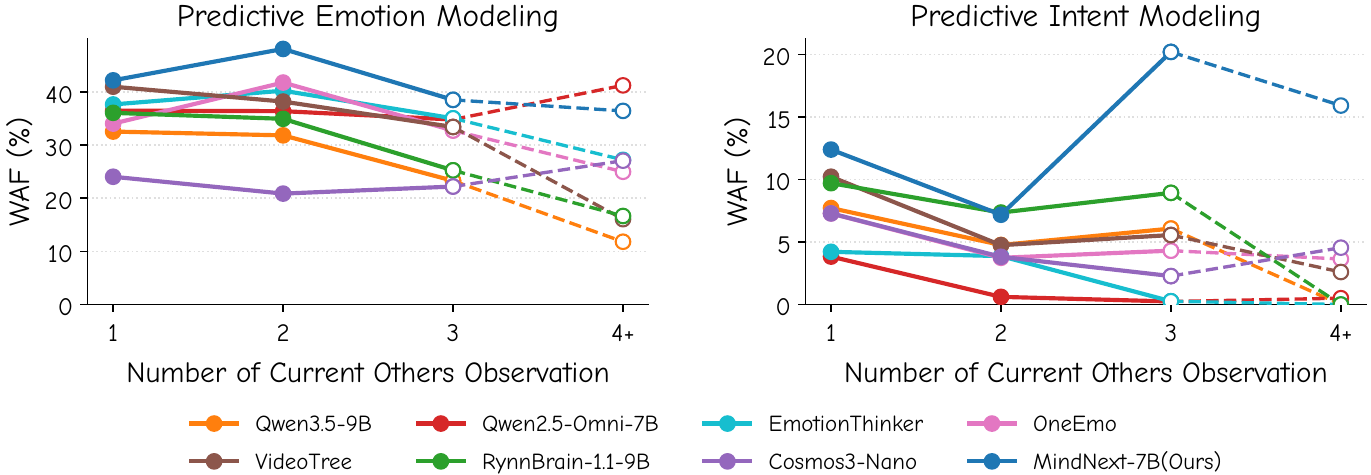}
    \caption{\textbf{Impact of current interacting peer scale on predictive mental modeling.} Performance comparison on predictive emotion (left) and intent (right) modeling across varying numbers of observed peers. Dashed lines indicate the $4+$ peer long-tail group.}
    \label{fig:other_obs_influence}
\end{figure}

\begin{figure*}[t]
    \centering
    \includegraphics[width=1.0\linewidth]{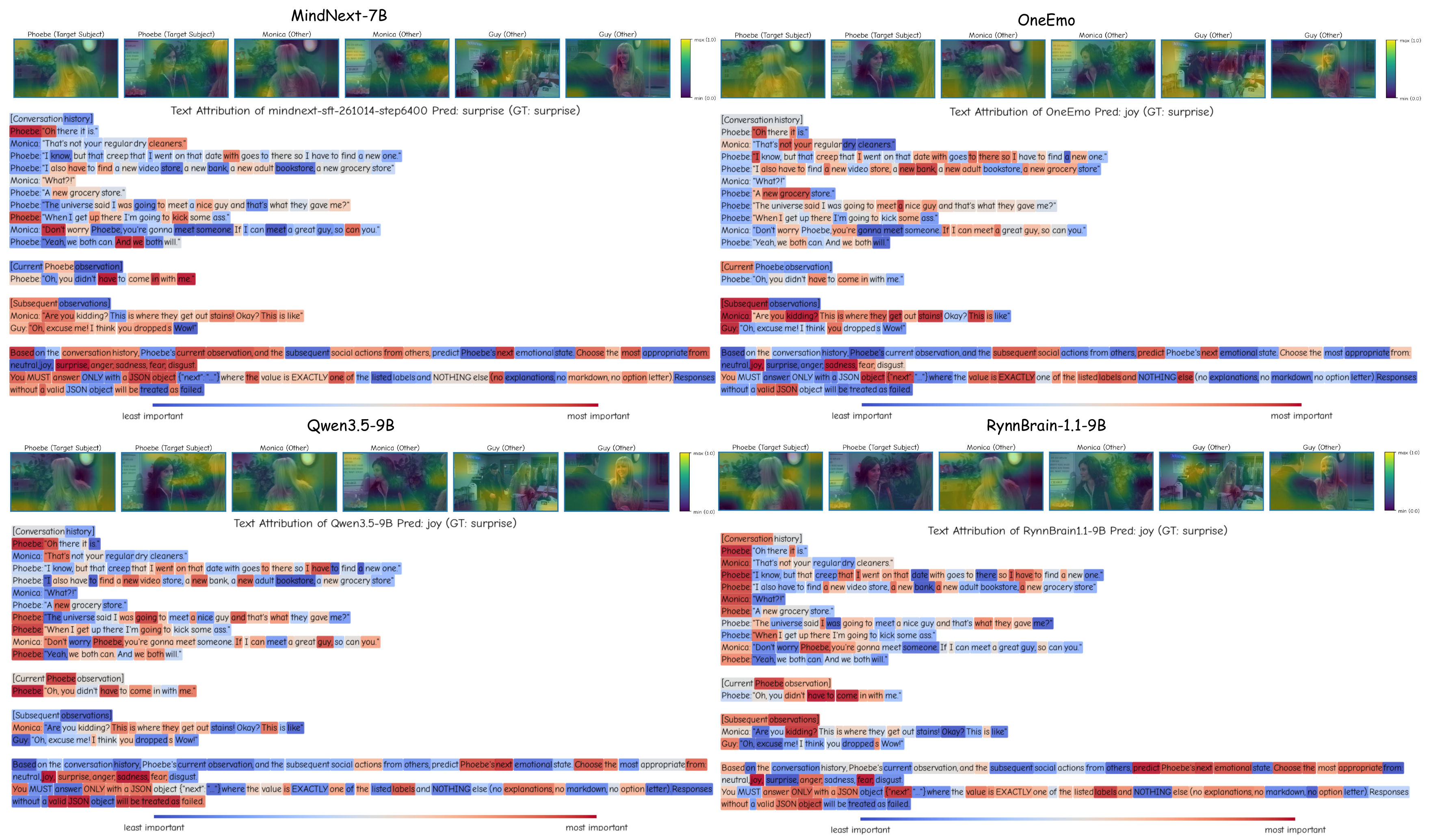}
    \caption{\textbf{Word-level and visual attribution analysis on \benchname.} Our model depends on sparser and more concentrated visual cues, ultimately predicting the correct result.}
    \label{fig:attribution-appendix}
\end{figure*}

\noindent{\textbf{Future Mental State Prediction Remains Sensitive to Context Density.}} 
As illustrated in \Cref{fig:other_obs_influence}, the performance of most baseline models undergoes an overall degradation as the number of observed interacting peers increases, exposing their fundamental fragility in effectively leveraging multi-party social actions. This bottleneck might stem from the inherent complexity of modeling group dynamics: as more interactants participate in the exchange, they contribute diverse viewpoints and exert compounded social influences that continuously perturb the target subject's evolving beliefs, emotions, and intentions.

\begin{table}[t]
\centering
\caption{Human evaluation of predictive behavioral modeling on 100 stratified test
samples.}
\label{tab:human_eval}
\small
\setlength{\tabcolsep}{5pt}
\begin{tabular}{lccc}
\toprule
Model & Overall $\uparrow$ & Verbal $\uparrow$ & Nonverbal  $\uparrow$ \\
\midrule
MindNext-7B     & \textbf{44.9} & \textbf{38.2} & \textbf{51.6} \\
VideoARM       & 40.8 & 35.7 & 46.0 \\
GLM-4.6V-Flash & 38.6 & 35.1 & 42.1 \\
Qwen3.5-9B     & 37.3 & 34.7 & 39.9 \\
Cosmos3-Nano   & 35.3 & 32.1 & 38.5 \\
RynnBrain1.1-9B   & 32.8 & 31.0 & 34.6\\
OneEmo         & 28.3 & 25.9 & 30.8\\
EmotionThinker & 19.4 & 29.0 & 9.9\\
\bottomrule
\end{tabular}
\end{table}

\noindent\textbf{Human Evaluation of Future Behavioral Modeling.}
Table 1 presents the human evaluation results across eight models on 100 stratified test samples. MindNext-7B achieves the highest performance across all three dimensions, scoring 44.9 Overall, 38.2 Verbal, and 51.6 Nonverbal. Compared with the runner-up VideoARM at 40.8, 35.7, and 46.0 on the same metrics, MindNext-7B leads by 4.1, 2.5, and 5.6 points, respectively. The remaining baselines follow with Overall scores ranging from 38.6 for GLM-4.6V-Flash down to 19.4 for EmotionThinker. To assess annotation reliability, we measure inter-annotator agreement between the two evaluators, obtaining a Spearman's $\rho$ of 0.857. Furthermore, human evaluation exhibits strong rank correlation with the automatic metrics. Specifically, human scores yield a Spearman's $\rho$ of 0.842 against LLM-Judge with a 95\% confidence interval of 0.820 to 0.861, and 0.816 against BSS with a 95\% confidence interval of 0.791 to 0.838. Overall, this alignment indicates high consistency in model rankings between human judgment and LLM-Judge, supporting the use of LLM-Judge for automated model comparisons on the test set.

\noindent\textbf{Visual Evidence Can Sometimes Mislead Models.}
As illustrated in \Cref{fig:attribution-appendix}, we visualize the attribution weights when predicting Phoebe’s future emotional state, where MindNext-7B correctly predicts surprise, while the three baseline models incorrectly predict joy. The visualization reveals that the baselines predominantly attend to Phoebe’s smiling expression during her final-frame interaction. In contrast, MindNext-7B avoids being biased by this superficial facial region.
At the word-level, while all models successfully attend to task constraints (\eg, future state prediction and formatting guidelines) and focus on candidate labels within the emotion space, their reliance on contextual evidence diverges. Crucially, MindNext-7B prioritizes verbal cues from interlocutors over Phoebe's visual facial expression. This case study demonstrates that the ability to disentangle decisive context from distracting superficial clues is essential for accurate future emotion forecasting.

\begin{figure}[ht]
    \centering
    \includegraphics[width=1.0\linewidth]{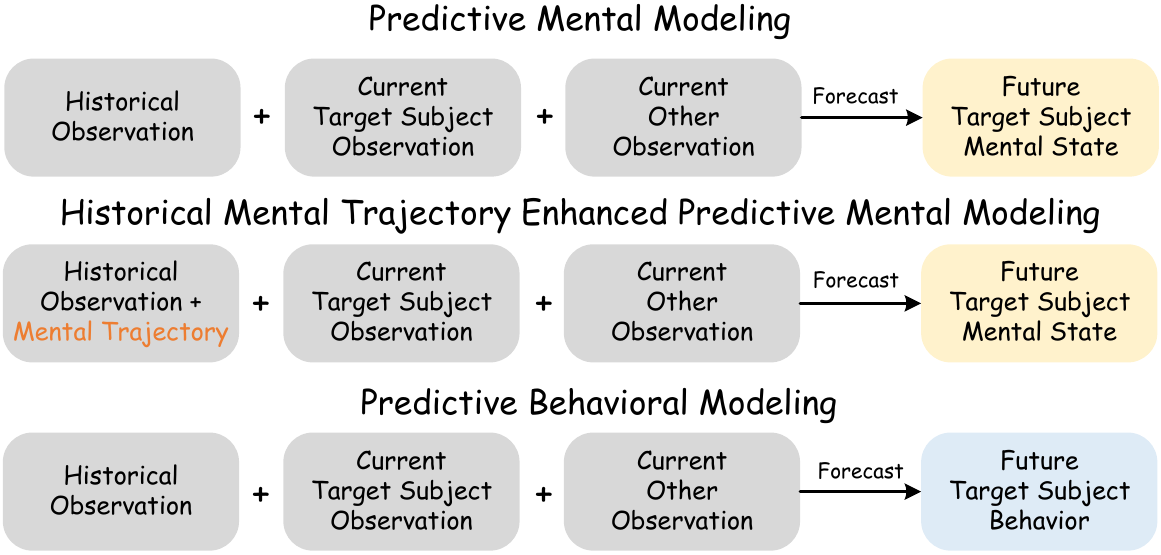}
    \caption{Overview of three different context and task outputs.}
    \label{fig:context}
\end{figure}
\section{Prompt Protocol}
\label{supsec:prompts}
We present the task prompts used in M$^3$WME-23K and M$^3$WME-Bench. \Cref{fig:context} illustrates three prompt types, including predictive mental modeling, historical-trajectory-enhanced mental modeling, and predictive behavioral modeling. The experiments for the trajectory-enhanced setting are reported in \Cref{supsec:additional-experiments}. We also include the evaluation prompt used for the LLM-Judge rubrics.

\begin{tcolorbox}[
  colback=gray!5, colframe=gray!60,
  title=\textbf{Predictive Mental Modeling Task Prompt},
  fonttitle=\small, fontupper=\ttfamily\footnotesize,
  breakable, enhanced,
  pad at break*=2mm,      
  left=2pt, right=2pt, top=2pt, bottom=2pt
]
[Conversation history]\\
<Speaker>: "<Utterance>"\\
<Target Subject>: "<Utterance>"\\
...\\

[Current <Target Subject> observation]\\
<Target Subject>: "<Utterance>"\\
<video><audio>\\

[Subsequent <Speaker> observation]\\
<Speaker>: "<Utterance>"\\
<video><audio>\\
...\\

Based on the conversation history, {TARGET}'s current observation, and the subsequent
social actions from others, predict {TARGET}'s next intention. Choose the most
appropriate from: [Mental Space List].

You MUST answer ONLY with a JSON object {"next": "..."} where the value is EXACTLY one
of the listed labels and NOTHING else (no explanations, no markdown, no option letter).
Responses without a valid JSON object will be treated as failed.
\end{tcolorbox}

\begin{tcolorbox}[
  colback=gray!5, colframe=gray!60,
  title=\textbf{Historical Mental Trajectory Enhanced PMM Prompt},
  fonttitle=\small, fontupper=\ttfamily\footnotesize,
  breakable, enhanced,
  pad at break*=2mm,      
  left=2pt, right=2pt, top=2pt, bottom=2pt
]
[Conversation history]\\
<Speaker>: "<Utterance>"\\
<Target Subject>: "<Utterance>"\\
...\\

[<Target Subject> (Emotion|Intent) history]\\
<Target Subject>: <mental state>\\
...\\

[Current <Target Subject> observation]\\
<Target Subject>: "<Utterance>"\\
<video><audio>\\

[Subsequent <Speaker> observation]\\
<Speaker>: "<Utterance>"\\
<video><audio>\\
...\\

Based on the conversation history, {TARGET}'s current observation, and the subsequent
social actions from others, predict {TARGET}'s next intention. Choose the most
appropriate from: [Mental Space List].

You MUST answer ONLY with a JSON object {"next": "..."} where the value is EXACTLY one
of the listed labels and NOTHING else (no explanations, no markdown, no option letter).
Responses without a valid JSON object will be treated as failed.
\end{tcolorbox}
\begin{tcolorbox}[
  colback=gray!5, colframe=gray!60,
  title=\textbf{Predictive Behavioral Modeling Task Prompt},
  fonttitle=\small, fontupper=\ttfamily\footnotesize,
  breakable, enhanced,
  pad at break*=2mm,      
  left=2pt, right=2pt, top=2pt, bottom=2pt
]
[Conversation history]\\
<Speaker>: "<Utterance>"\\
<Target Subject>: "<Utterance>"\\
...\\

[Current <Target Subject> observation]\\
<Target Subject>: "<Utterance>"\\
<video><audio>\\

[Subsequent <Speaker> observation]\\
<Speaker>: "<Utterance>"\\
<video><audio>\\
...\\

Based on {Target Subject}'s inferred current mental state, and imagining how the subsequent
social actions may implicitly reshape {Target Subject}'s plausible future mental state,
predict {Target Subject}'s next verbal and nonverbal action.

You MUST answer ONLY with a JSON object:\\
{\\
  "verbal": "{TARGET}'s next utterance",\\
  "nonverbal": [\\
    "face: expression, gaze etc.",\\
    "gesture: hand, head motion etc.",\\
    "body: <posture, body action etc>"\\
  ]\\
}\\

Rules:\\
1. "verbal" = {TARGET}'s next utterance only. \\
2. "nonverbal" = only {TARGET}'s visible
next behavior. \\
3. Use only face/gesture/body.\\ 
4. JSON format output only.\\
\end{tcolorbox}
\begin{tcolorbox}[
  colback=gray!5, colframe=gray!60,
  title=\textbf{The LLM Scoring Rubrics},
  fonttitle=\small, fontupper=\ttfamily\footnotesize,
  breakable, enhanced,
  pad at break*=2mm,      
  left=2pt, right=2pt, top=2pt, bottom=2pt
]
You are a strict judge of one person's NEXT action.
Compare PRED with GT only. Score observable action, not wording, length, or position.\\

Score fields separately: verbal, and nonverbal (face, gesture, body).
Allowed scores: 0, 20, 40, 60, 80, 100.\\

Score rubric:
- 100 Equivalent: same action; paraphrase is fine.\\
- 80 Minor gap: same action; only a non-core detail is missing or added.\\\\
- 60 Partial: same action family, but a core attribute is wrong or missing.\\
- 40 Mixed: some overlap, yet the described action is incomplete or different.\\
- 20 Weak: only a faint related cue; the main action is wrong.\\
- 0 Incorrect: unrelated, contradictory, empty vs non-empty, or wrong slot.\\

Core vs non-core:
- Read each field as atomic actions (split on comma/semicolon):\\
- Action type: utterance meaning / smile / look / shrug / nod / point / lean / stand / turn / leave.\\
- Core attribute: gaze target, point target, body direction, utterance meaning.\\
- Modifier: intensity or manner. It never moves a score except 100 vs 80.\\
- verbal core = utterance meaning.\\
- face core = expression type and/or gaze target.\\
- gesture core = gesture type (shrug, nod, point, wave...).\\
- body core = body-action type (lean, stand, sit, turn, leave...).\\
- Non-core = intensity/degree/laterality (slight/small, left/right) and extra modifiers.\\

Decision rules:\\
1. Both empty = 100. One empty = 0.\\
2. Score each field against the same field only. Do not move content across slots. Wrong-slot content = 0 there; the true GT field is still missing.\\
3. Same action types and same core attributes = 100, or 80 if only modifiers differ.\\
4. Action type matches, one core attribute does not = 60. If that core attribute is the whole field, = 0.
   Example: "looks at Monica" vs "looks at Rachel" = 0.
   Example: "tight smile, looks at Monica" vs "tight smile, looks at Rachel" = 60.\\
5. Missing some GT atomic actions, or adding a different extra action = 40.\\
6. No matched action type, only a vague related cue = 20.
\end{tcolorbox}
